\documentclass[a4paper,11pt]{article}
\usepackage{url}
\usepackage{graphicx}
\usepackage[english]{babel}
\usepackage{cite}
\usepackage{amsmath,amsthm}
\usepackage{amsfonts}

\usepackage{arxiv}

\title{rfPhen2Gen: A machine learning based association study of brain imaging phenotypes to genotypes}

\author{Muhammad Ammar Malik\\
Department of Informatics\\
University of Bergen\\
PO Box 7803, 5020 Bergen, Norway\\
\texttt{muhammad.malik@uib.no}\\
\And
Alexander Lundervold\\
Department of Computer Science\\
Western Norway University of Applied Sciences\\
PO Box 7030, 5020 Bergen, Norway\\
\texttt{alexander.selvikvag.lundervold@hvl.no}\\
\And
Tom Michoel\\
Department of Informatics\\
University of Bergen\\
PO Box 7803, 5020 Bergen, Norway\\
\texttt{tom.michoel@uib.no}\\
}

\renewcommand{\shorttitle}{\textit{arXiv} Template}

\begin{document}
\maketitle
\begin{abstract}
Imaging genetic studies aim to find associations between genetic variants and imaging quantitative traits. Traditional genome-wide association studies (GWAS) are based on univariate statistical tests, but when multiple traits are analyzed together they suffer from a multiple-testing problem and from not taking into account correlations among the traits. An alternative approach to multi-trait GWAS is to reverse the functional relation between genotypes and traits, by fitting a multivariate regression model to predict genotypes from multiple traits simultaneously. However, current reverse genotype prediction approaches are mostly based on linear models. Here, we evaluated random forest regression (RFR) as a method to predict SNPs from imaging QTs and identify biologically relevant associations. We learned machine learning models to predict 518,484 SNPs using 56 brain imaging QTs. We observed that genotype regression error is a better indicator of permutation p-value significance than genotype classification accuracy. SNPs within the known Alzheimer disease (AD) risk gene APOE had lowest RMSE for lasso and random forest, but not ridge regression. Moreover, random forests identified additional SNPs that were not prioritized by the linear models but are known to be associated with brain-related disorders. Feature selection identified well-known brain regions associated with AD, like the hippocampus and amygdala, as important predictors of the most significant SNPs. In summary, our results indicate that non-linear methods like random forests may offer additional insights into phenotype-genotype associations compared to traditional linear multi-variate GWAS methods.
\end{abstract}


\section{INTRODUCTION}
Neuroimaging genetics, also known as imaging genomics or imaging genetics, is a useful tool to investigate the associations between genetic variants and variation in brain structure among individuals \cite{shen2010whole}. The discovery of biomarkers jointly from imaging and genetic data helps us to better understand the underlying pathological processes of neuropsychiatric and neurodegenerative diseases \cite{chauhan2015association, bi2017genome}. Moreover, neuroimaging may help us discover the genetic pathways through which genes affect the above-mentioned diseases, by identifying associations between causal genes and variations in brain regions \cite{lu2017bayesian, liu2014cardiovascular}.  And lastly, imaging genetics studies have been shown to have increased statistical power when compared with conventional case-control studies and therefore have decreased sample size requirement \cite{potkin2009genome}.


Recently a large number of neuroimaging studies have been conducted to explore the association between neurodegenerative disease and brain structure \cite{wang2012identifying, wang2012phenotype,shen2010whole, huang2015fvgwas, huang2019spatial}. Some of these studies have focused on understanding the genetic causes of these diseases  (for example Alzheimer’s disease), whereas the other genome-wide association studies (GWAS) focus on identifying the genetic variations that influence brain structure and function. A common issue with most imaging genetics studies is the reduction in either imaging or genetic data (or sometimes both). For example, whole-brain studies have mostly focused on a small number of genetic variants \cite{zhou2018brain,  hariri2006imaging, brun2009mapping, shen2007morphometric}, whereas whole-genome studies have focused on a limited number of imaging quantitative traits (QTs) \cite{potkin2009hippocampal, baranzini2009genome}. This restriction in either genotype or phenotype data can greatly hinder our ability to identify important associations. 

A typical procedure to investigate genotype-phenotype associations is to conduct univariate linear regression or analysis of variance (ANOVA) tests for each genetic variant against each trait separately. However, when multiple traits are studied simultaneously this approach ignores correlations among traits and leads to a high multiple-testing burden. Other approaches for multi-trait GWAS are based on multivariate analysis of variance (MANOVA) or canonical correlation analysis (CCA)\cite{ferreira2009multivariate}. But these are applicable only to studies with a small number of traits. A promising alternative approach to multi-trait GWAS has been to reverse the functional relation between genotypes and traits and fit a multivariate regression model that predicts genotypes from multiple traits simultaneously, instead of the usual approach to regress traits on genotypes \cite{o2012multiphen}.  In fact, in \cite{malik2021high}, we showed that more traditional multi-trait GWAS methods such as CCA can also be described as reverse genotype prediction methods.

Reverse genotype prediction has also been considered in the context of imaging genetics. For example, in \cite{wang2012phenotype} a task-correlated longitudinal sparse regression approach was used to investigate associations between phenotype markers and Alzheimer’s disease relevant SNPs belonging to top 40 AD relevant genes. 

Thus far, efforts to extend reverse genotype prediction methods to the high-dimensional setting (in either SNP or feature dimension) have mostly focused on gene expression traits. For instance, a recent study used L2-regularized linear regression of SNPs on gene expression traits to identify trans-acting expression quantitative trait loci (trans-eQTLs), and showed that this approach aggregates evidence from many small trans-effects while being unaffected by strong expression correlations \cite{banerjee2021tejaas}.

\begin{figure*}[b!]
    \centering
    \includegraphics[width=1.0\linewidth]{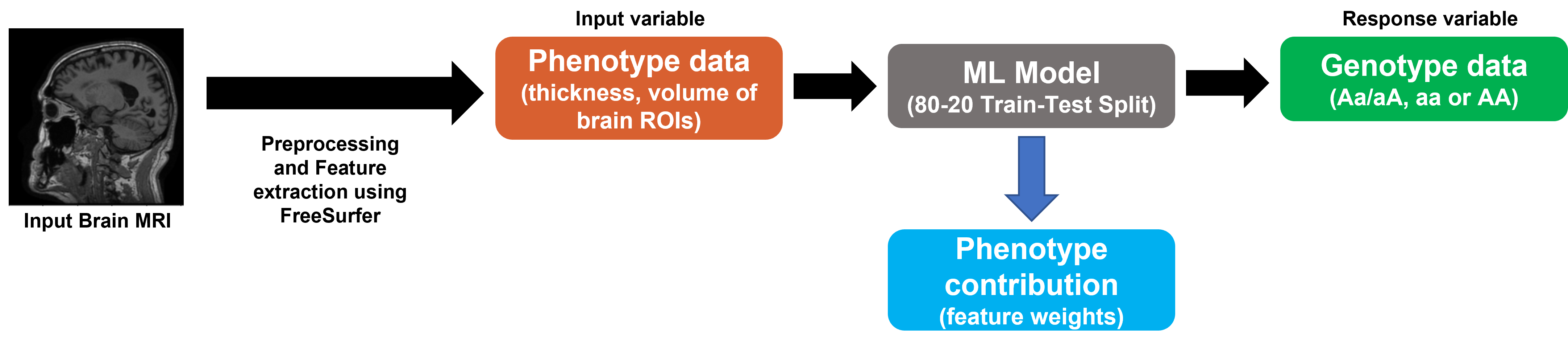}
    \caption{Flowchart of the approach used in this study}
    \label{fig:approach}
\end{figure*}

However, several limitations and open questions remain in multi-trait GWAS. For example, existing studies mostly rely on linear models that search for linear combinations of traits associated to the SNP, but there is no \textit{a priori} biological evidence to support the use of only linear combinations. Moreover, even though L2-regularization allows to deal with high-dimensional traits, it does not address the problem of feature selection. For instance, in \cite{banerjee2021tejaas} a secondary set of univariate tests is carried out to select genes associated to trans-eQTLs identified by the initial multivariate regression. 

In \cite{malik2021high}, using gene expression data  from a cross between two yeast strains, we found that feature coefficients of machine learning models (lasso, ridge, linear SVM, and random forest) correlated with the strength of association between variants and individual traits, and were predictive of true trans-eQTL target genes. However, to the best of our knowledge, a genome-wide analysis of machine learning methods for reverse genotype prediction in human GWAS has not yet been conducted.

In this study, we explored the use of a non-linear machine learning method, specifically random forests, for predicting genotypes from brain imaging phenotypes. Our overall hypothesis is that genetic variants whose genotypes can be predicted with higher accuracy from imaging traits are more likely to affect some or all of the traits under consideration than variants whose genotypes cannot be predicted well, and that feature weights in the fitted models measure the strength of biological association between a SNP and a QT. We compare the results of random forest with two well-known linear methods, lasso and ridge regression. 

\section{MATERIALS AND METHODS}

\subsection{Dataset}
Data used in this study was obtained from the ADNI database (\url{adni.loni.ucla.edu}). The baseline T1-weighted MRI images from the four phases of the ADNI study, the Illumina SNP genotyping data, demographic information, APOE genotype, and baseline diagnosis was downloaded from the ADNI database. The demographic information of the samples used in the study can be found in Table \ref{tab:demo}. Details about the standardized imaging protocols used in ADNI can be found in \url{https://adni.loni.usc.edu/methods/documents/mri-protocols/}. 

\begin{table}[h!]
\caption{Demographic information of the samples used in the study. CN: Controls, MCI: Mild Cognitive Impairment, AD: Alzheimer's Disease}
    \begin{center}
    \begin{tabular}{|c|c|c|c|}
    \hline
    \textbf{}&\textbf{CN}&\textbf{MCI}&\textbf{AD}\\
    \hline
    No. of subjects & 211 & 359 & 178 \\ \hline
    Gender(M/F) & 112/99 & 234/125 & 94/84 \\ \hline
    Baseline age(years:mean $\pm$ SD) & 75.7$\pm$4.9 & 74.7$\pm$7.3 & 75.4$\pm$7.3 \\ \hline
	Education(years:mean $\pm$ SD) & 16.0$\pm$2.8 & 15.7$\pm$3.0 & 14.6$\pm$3.2 \\ \hline
	Race (Caucasian/Non-Caucasian) & 191/20 & 325/34 & 161/17 \\ \hline
    \end{tabular}
        \label{tab:demo}    
    \end{center}
\end{table}

\subsection{MRI data and imaging phenotype extraction}
We extracted subcortical segmentation and cortical parcellation from the T1-weighted images using FreeSurfer v6.0~\cite{fischl2012freesurfer} to obtain imaging phenotypes. Following \cite{shen2010whole} we defined 56 volumetric and cortical thickness values mentioned in(Table \ref{tab:pheno_list}).

\begin{table}[t!]
    \caption{List of FreeSurfer phenotypes defined as volume or cortical thickness of various region of interests (ROI)$^\mathrm{a}$}
    \centering
    \begin{tabular}{|p{8cm}|p{8cm}|}
    \hline
    \multicolumn{2}{|c|}{\textbf{Phenotype description (Phenotype ID)}}\\\hline
    Volume of amygdala (AmygVol) & Volume of cerebral cortex (CerebCtx) \\ \hline
    Volume of cerebral white matter (CerebWM) & Volume of hippocampus (HippVol) \\ \hline
    Volume of inferior lateral ventricle (InfLatVent) & Volume of lateral ventricle (LatVent) \\ \hline
    Thickness of entorhinal cotrex (EntCtx) & Thickness of fusiform gyrus (Fusiform) \\ \hline
    Thickness of inferior parietal gyrus (InfParietal) & Thickness of inferior temporal gyrus (InfTemporal) \\ \hline
    Thickness of middle temporal gyrus (MidTemporal) & Thickness of parahippocampal gyrus (Parahipp) \\ \hline
    Thickness of posterior cingulate (PostCing) & Thickness of postcentral gyrus (Postcentral) \\ \hline
    Thickness of precentral gyurs (Precentral) & Thickness of precuneus (Precuneus) \\ \hline
    Thickness of superior frontal gyrus (SupFrontal) & Thickness of superior parietal gyurs (SupParietal) \\ \hline
    Thickness of superior temporal gyrus (SupTemporal) & Thickness of supramarginal gyrus (Supramarg) \\ \hline
    Thickness of temporal pole (TemporalPole) & \\ \hline
    \multicolumn{2}{|p{16cm}|}{Mean thickness of caudal anterior cingulate,
isthmus cingulate, posterior cingulate, and
rostral anterior cingulate (MeanCing)} \\ \hline
    \multicolumn{2}{|p{16cm}|}{Mean thickness of caudal midfrontal, rostral
midfrontal, superior frontal, lateral orbitofrontal,
and medial orbitofrontal gyri and frontal pole (MeanFront)} \\ \hline
    \multicolumn{2}{|p{16cm}|}{Mean thickness of inferior temporal, middle temporal,
and superior temporal gyri (MeanLatTemp)} \\ \hline
    \multicolumn{2}{|p{16cm}|}{Mean thickness of fusiform, parahippocampal,
and lingual gyri, temporal pole and transverse
temporal pole (MeanMedTemp)} \\ \hline
    \multicolumn{2}{|p{16cm}|}{Mean thickness of inferior and superior parietal gyri,
supramarginal gyrus, and precuneus (MeanPar)} \\ \hline
    \multicolumn{2}{|p{16cm}|}{Mean thickness of precentral and postcentral gyri (MeanSensMotor)} \\ \hline
    \multicolumn{2}{|p{16cm}|}{Mean thickness of inferior temporal, middle temporal,
superior temporal, fusiform, parahippocampal , and
lingual gyri, temporal pole and transverse temporal pole (MeanTemp)} \\ \hline
\multicolumn{2}{p{16cm}}{$^{\mathrm{a}}$Each of the 28 phenotypes mentioned corresponds to two phenotypes, one for the left side and the other for the right side.}    
    \end{tabular}
        \label{tab:pheno_list}
\end{table}

\subsection{SNP genotypes}

The SNP data from ADNI database were genotyped using the Human 610-Quad BeadChip (Illumina, Inc., San Diego, CA, USA). The genotype data consists of 620,901 SNPs. The SNP data was screened using the following quality control (QC) steps: (1) call rate check per subject ($\geq 90\%$) and per SNP marker ($\geq 90\%$), (2) gender check (3) marker removal according to the minor allele frequency (MAF) $\geq 5\%$ and (4) Hardy-Weingberg equilibrium (HWE) test of $p\leq10^{-6}$. The remaining missing genotype values were imputed as the modal values. After the QC procedure, 749 subjects and 518,484 SNPs remained in the data. The
APOE gene is one of the important causal genes for AD, but the previously identified APOE SNPs (rs429358/rs7412) were not available on the Illumina array. Therefore, the APOE genotype was coded from the \texttt{ADNIMERGE.csv} file prepared by the ADNI study by using the number of APOE-{$\varepsilon$}4 risk alleles.

\subsection{Genotype prediction model}
For genotype prediction using machine learning models, the phenotype values were treated as explanatory variables whereas the genotype values of SNPs were treated as the response variables, learning separate models for each SNP (Fig. \ref{fig:approach}). We used both regression and classification to predict SNPs. The prediction performance in the regression setting was measured by computing the root mean squared error (RMSE) between the predicted and the actual genotype value of a variant. The classification performance was measured as the ratio between the number of correctly predicted samples and the total number of samples in the test set.

We compared the performance of random forest regression and classification (RFR), with linear machine learning methods, in particular  ridge regression (RR) and lasso regression (LR).

\subsection{Phenotype contribution}
We used the feature weights of the machine learning models for measuring the association between SNPs and specific brain regions (in the case of RFR feature importances were used). The absolute values of the feature weights were used and normalized such that they sum up to 1.

\subsection{Permutation tests}

We conducted permutation tests \cite{edgington2007randomization} using 100 permutations for a subset of 876 SNPs to determine the statistical significance of RMSE and classification accuracies. \textit{P}-values were calculated as the fraction of permutation values that are at least as extreme as the original statistic (RMSE or classification accuracy) derived from non-permuted data, that is
\begin{equation}
    p = \frac{i+1}{N}
    \label{eq:perm_p}
\end{equation}
where $N$ denotes the number of permutations, and $i$ denotes the number of times the performance measure  of the permuted SNP was found to be better than the unpermuted measure for that SNP.

\subsection{Rank-based p-values}

Because conducting permutation tests for all 518,484 SNPs was computationally prohibitive, we computed the p-values for all the SNPs in the genome by ranking the SNPs based on the obtained RMSE values (for tied ranks the average of the rank was assigned) and dividing the rank by the total number of SNPs (Eq. \ref{eq:p_whole}).

\begin{equation}
    p = \frac{k}{N}
    \label{eq:p_whole}
\end{equation}
where $N$ denotes the total number of SNPs, and $k$ denotes the rank of the particular SNP. These rank-based p-values for visualizing relative prediction performance of SNPs in Manhattan plots.

\subsection{Experimental settings}
The machine learning models including lasso regression, ridge regression and random forest regression and classification were implemented using the \textit{scikit-learn} Python library. 
The dataset was divided into 80-20 training-test split. Both the input and output data were normalized to have unit standard deviation and zero mean. The volumes of the ROIs (phenotype data) were corrected for age, gender, education, and baseline Intracranial Volume (ICV$\_$bl) as estimated by FreeSurfer v6.0.

\begin{figure}[b!]

\begin{minipage}[t]{0.50\linewidth}
  \textbf{A}\\
      \includegraphics[width=\linewidth]{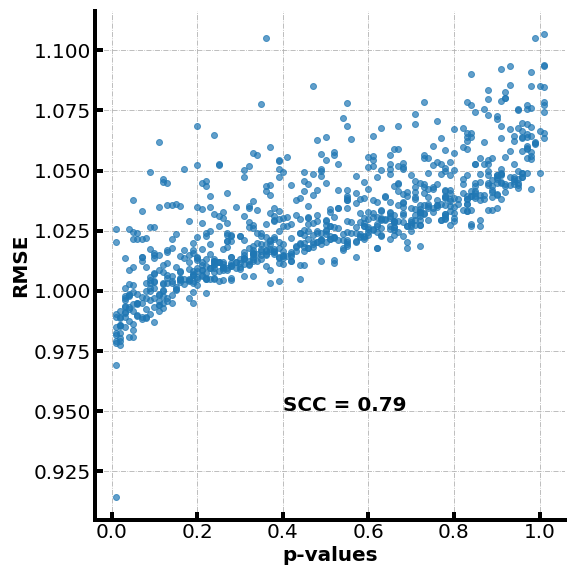}
      
  \end{minipage}
  \hfill
  \begin{minipage}[t]{0.50\linewidth}
  \textbf{B}\\
      \includegraphics[width=\linewidth]{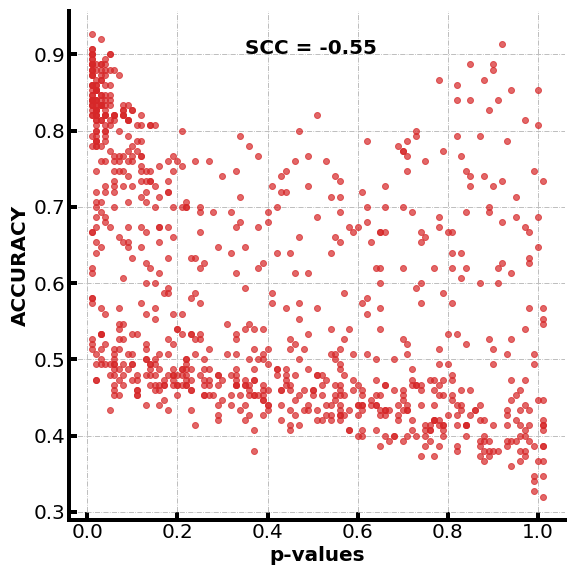}
\end{minipage}

  \caption{Permutation tests for Random Forest. \textbf{(A)} Regression (RMSE), \textbf{(B)} Classification (Accuracy). SCC: Spearman Correlation Coefficient} 
  \label{fig:perm_tests}
\end{figure}

\section{Results}

\subsection{Genotype classification vs genotype regression}

SNP genotypes take discrete values (0,1,2) counting the number of alternative alleles in an individual. Since it is generally assumed that the effect of a SNP on a quantitative trait increases or decreases with the number of alternative alleles, genotype prediction could either use classification models (emphasizing the discrete nature) or regression models (emphasizing the ordinal nature). To compare and select between regression or classification for the genotype prediction task, we performed permutation tests to convert prediction performance  (\textit{root mean squared error} in case of regression and \textit{accuracy} in case of classification) to p-values. 

Since performing permutation tests across the whole genome was computationally unfeasible, we considered 876 SNPs belonging to the top 40 AD-related genes as mentioned on \url{alzgene.org}. For the permutation tests, the target labels were randomly permuted 100 times and the results were compared with unpermuted data. Based on eq. \ref{eq:perm_p} the best possible p-value in this scenario is 0.01, i.e. none of the permuted sets performed better than the unpermuted set.

The correlation between \textit{p-values} and the prediction task performance was found to be higher for regression (\textbf{Spearman correlation coefficient 0.79}) than for classification  (\textbf{Spearman correlation coefficient -0.55})
(Fig. \ref{fig:perm_tests}). Further investigation showed that the poor correspondence between classification accuracy and p-values was due to strong class imbalance (Supp. Fig. S1): if alternative alleles of a SNP are relatively rare (few individuals in the 1 and/or 2 class), classification accuracy can be high by randomly assigning individuals to classes based on the class frequencies such that classification accuracy from real features is no better than from random features. In contrast, as indicated in  Fig. \ref{fig:perm_tests}, genotype regression was less affected by class imbalance differences among SNPs.

Since RMSE is a better indicator of non-random prediction performance, regardless of minor allele frequency differences between SNPs, than classification accuracy, we decided to proceed with regression analysis for the remainder of our experiments.

\subsection{Genotype prediction across the whole genome}
To the best of our knowledge, none of the previous reverse genotype prediction studies were conducted across the whole human genome. We performed reverse genotype regression for all 518,484 SNPs that passed QC using Lasso, Ridge, and Random Forest regression on 56 volumetric and cortical thickness image features. For visualization purposes, we converted the RMSE value to a rank-based p-value for each SNP (eq. \ref{eq:p_whole}). 

\begin{figure*}[t!]
    \centering
    \includegraphics[width=\linewidth]{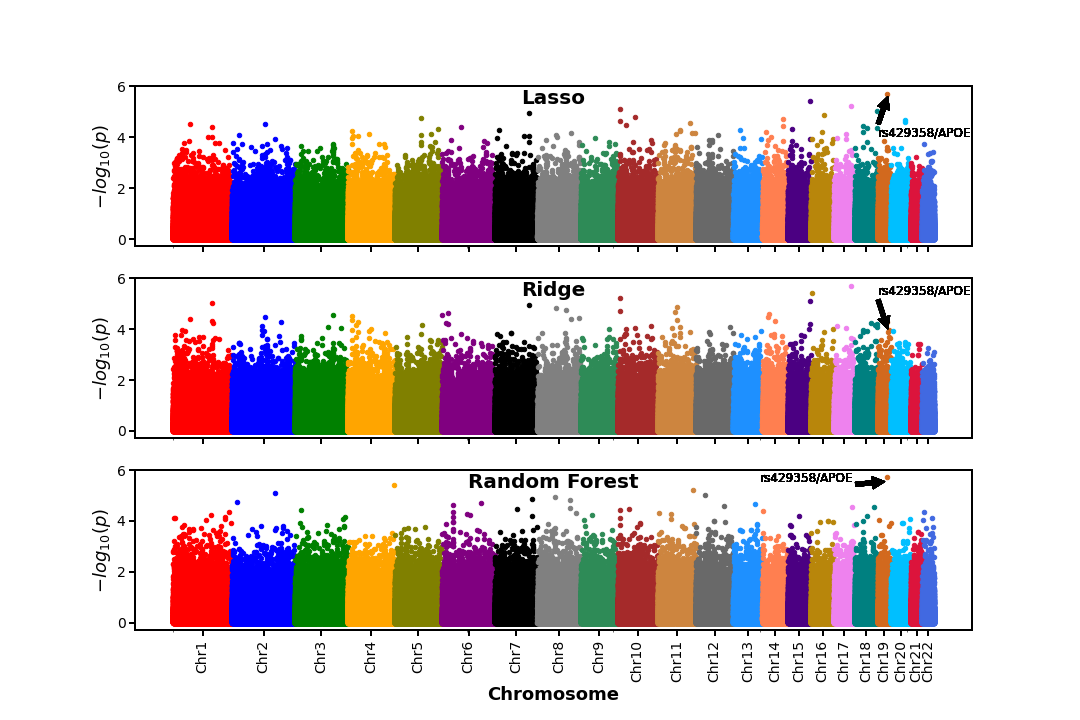}
    \caption{Negative log transformed p-values plotted across the whole genome for Random Forest Regression (RFR), Ridge Regression (RR), Lasso Regression (LR).}
    \label{fig:manhattan_all}
\end{figure*}

In support of our overall hypothesis, we observe that genotype prediction performance is variable across the genome and that some SNPs can be predicted with lower RMSE than the random background  (Fig. \ref{fig:manhattan_all}). Moreover, the top SNP achieving the lowest RMSE value for random forests (RMSE=0.915) and lasso (RMSE=0.932) was rs429358/APOE, the best known causal gene for Alzheimer's disease. In contrast, ridge regression picked up rs1864685/SLC39A11 as the best SNP (RMSE=0.926), with rs429358/APOE being ranked 27th (RMSE=0.950). 

Despite this difference on the APOE prediction, overall the linear methods (lasso and ridge) produced nearly the same ranking of SNPs (Spearman correlation coefficient 0.91), whereas random forests predictions were clearly distinct from the linear ones (Spearman correlation coefficients 0.25 and 0.29 with ridge and lasso, respectively) (Supp. Fig. S2). Different peaks across the methods further illustrate point to the fact that distinct SNPs are identified using different methods, also near the top of the ranking (Supp. Fig. S3).

To test whether any of these results were affected by population structure in the data, we repeated the analysis using only Caucasian subjects, and observed no significant difference in the RMSE distribution (Supp. Fig. S4).

Since lasso and ridge regression gave comparable results, and ridge regression has been well studied as a multivariate GWAS method \cite{banerjee2021tejaas}, we focused the remainder of our analysis on the random forest predictions. 

\subsection{Identification of genetically associated imaging markers}

\begin{figure*}[b!]
    \centering
    \includegraphics[width=\linewidth]{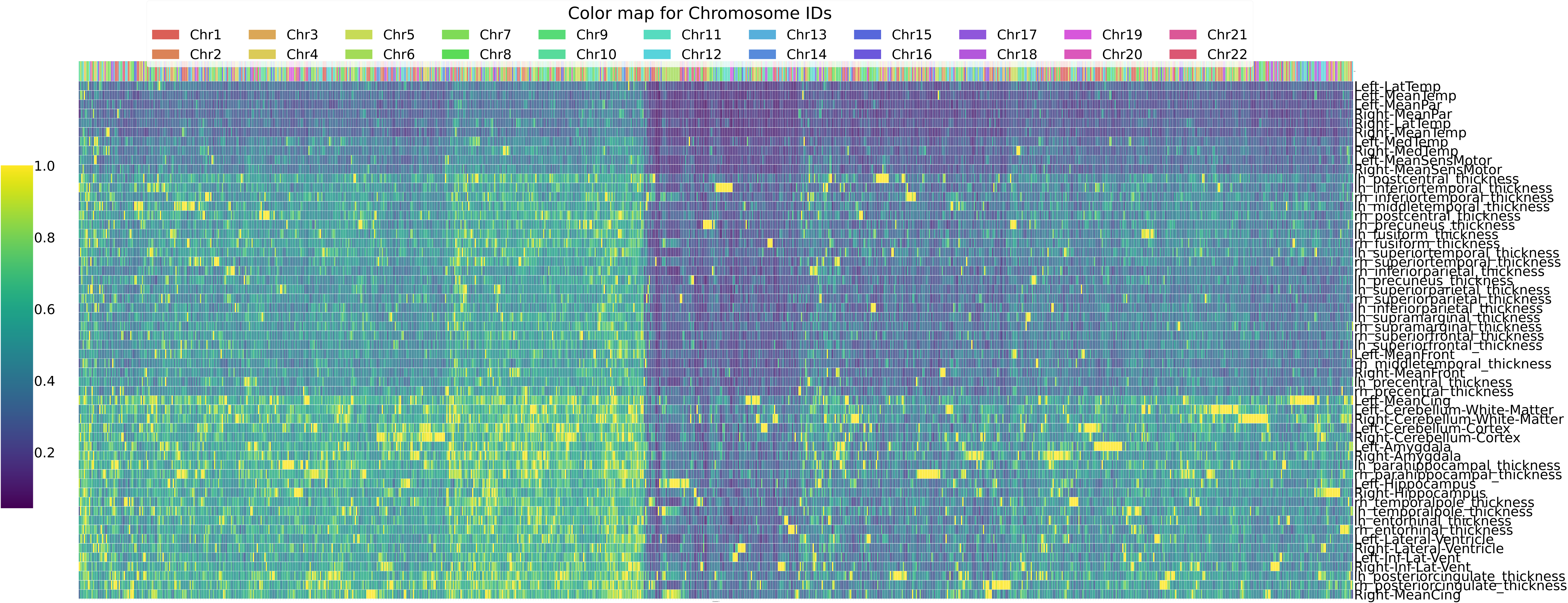}
    \caption{Heat maps of feature weights for top 1000 SNPs identified by random forest.}
    \label{fig:cmap}
\end{figure*}

One of the aims of this study is to test if a subset of genetic variants and imaging phenotypes related to brain-related disorders can be identified from the feature weights of the best performing SNPs. Figure \ref{fig:cmap} shows the clustermap of feature weights of all the imaging features and the top 1000 SNPs identified by the random forest method. A clear pattern can be noticed where groups of SNPs not colocated on the genome associate to similar groups of imaging features. This is consistent with previous findings from univariate approaches where genetic variants affecting one cortical region were found to often also affect other cortical regions \cite{shadrin2021vertex}. In particular, brain regions previously known to be associated with brain-related disorders, such as the  hippocampus, amygdala, temporal lobe, entorhinal cortex and lateral ventricles, clustered together.

The hippocampus has been shown to be related to memory and cognition \cite{squire2011cognitive}. The amygdala is known to play a primary role in decision making and memory \cite{amunts2005cytoarchitectonic}, and its atrophy is associated with Alzheimer's disease \cite{poulin2011amygdala}. The temporal lobe is  largely responsible for creating and preserving both conscious and long-term memory \cite{jack1998rate}. The temporal pole, also known as Broadman area 38,  is among the earliest affected by Alzheimer's disease, frontotemporal dementia and frontotemporal lobar degeneration \cite{arnold1991topographical}. The entorhinal cortex is one of the first regions to be affected in Alzheimer's disease \cite{khan2014molecular} and moreover, the lateral ventricles are known to be enlarged in patients with Alzheimer's \cite{nestor2008ventricular},  schizophrenia \cite{wright2000meta}, bipolar and major depressive disorders \cite{kempton2011structural}.

\subsection{Characterization of top genetic variants}

To further investigate the relevance of our findings, we analyzed
the top 10 SNPs that were identified by the random forest method in detail using literature search and data from the GWAS catalog \cite{macarthur2017new} (Table \ref{tab:topSNP}). As previously mentioned, the SNP rs429358 belonging to the APOE gene was identified as the top SNP. APOE genotype is the most well-known genetic risk factor for AD \cite{farrer1997effects}. The top 3 features identified for APOE are the volumes of the hippocampus, amygdala and the thickness of the entorhinal cortex, all of these are known to be among the first regions to be affected in Alzheimer's disease \cite{ amunts2005cytoarchitectonic, khan2014molecular}.

While, apart from APOE, the rest of the top SNPs do not belong to known genes associated with AD, all of them have been associated previously with other brain related functions (Table \ref{tab:topSNP}). For example, F11-AS1, SYBU, GACAT3, and LAMA2 are known to be associated with traits like cognitive decline, cortical thickness and surface area, and brain volume measurements \cite{van2020understanding, grasby2020genetic}. Moreover, FLI1 and TMEM213 have been shown to be associated with paired helical filaments (PHF)-tau measurements \cite{wang2020genome}. The tau protein is also a well-known Alzheimer's disease biomarker \cite{harrington1991measurement}.

\begin{table}[t]

\begin{center}
\caption{Top SNPs identified by random forest}
\begin{tabular}{|l|l|l|l|}
\hline
\textbf{SNP}        & \textbf{Gene}        & \textbf{Associated traits}                                                                                     & \textbf{Features identified}                                                      \\ \hline
rs429358   & APOE      & HDL cholesterol, LDL   cholesterol,                                  & Hippocampus, Amygdala,                                 \\
~		& ~						& Alzheimer's disease, Fatty liver & Entorhinal cortex  \\ \hline

rs2084729  & F11-AS1   & cognitive decline, cortical   thickness,              & Entorhinal cortex, Amygdala,                  \\
~		& ~						& cortical surface area, brain volume measurements		& Posterior cingulate cortex  \\ \hline

rs622735   & FLI1     & PHF-tau measurments                                                                                   & Parahippocampal gyrus, \\ 
~ & ~ & ~ & Superior   Temporal gyrus                         \\ \hline

rs3749030  & TTC21B    & intelligence, cognitive   function, measurements                                                      & MeanCing, Cerebellum White   Matter,                   \\
~       &   ~      &    ~ & Postcentral gyrus \\ \hline

rs11048593 & ITPR2     & HDL   cholesterol, total cholestrol,  & Cerebellum White Matter,                               \\
~		& ~						& LDL cholesterol levels, unipolar depression & MeanCing, Amygdala \\ \hline

rs6473635  & PXDNL      & cognitive impairment   measurement, unipolar depression,               & MeanCing, Temporal pole gyrus,                                   \\
~		& ~						& bipolar disorder, schizophrenia &	Amygdala \\ \hline

rs896189   & TMEM213  & PHF-tau measurments                                                                                   & inferior parietal gyrus,   postcentral gyrus,              \\ 

~           &   ~              & ~ &   fusiform gyrus \\ \hline

rs12678956 & SYBU      & Schizophrenia, cortical surface area                                                                  & Inferior lateral ventricle,    \\
~       &   ~                 &   ~   & posterior cingulate cortex, fusiform gyrus \\ \hline

rs6714955  & GACAT3      & cortical surface area   measurement,                       & posterior cingulate cortex,                   \\
~	&	~						& neuroimaging measurement, brain measurement	& Cerebellum Cortex, MeanCing  \\ \hline

rs9482965  & LAMA2     & cognitive ability, cortical   surface area,                                         & MeanTemp, posterior cingulate   cortex,       \\
~	&	~	 &  cortical thickness &	 Inferior lateral ventricle \\ \hline

\end{tabular}
\end{center}
\label{tab:topSNP}
\end{table}

\section{Discussion}
In this paper, we analyzed the use of machine learning methods for associating brain imaging phenotypes to genotypes, in a reverse genotype prediction setting, using data from 56 imaging quantitative traits and more than half a million SNPs. The basic hypotheses of reverse genotype prediction from multiple trait combinations are that variants whose genotypes can be predicted with higher accuracy are more likely to have a true effect on one or more of the measured traits, and that feature importances or coefficients in the trained models indicate the strength of association between variants and individual traits. However, existing studies only considered linear models. Here we performed a comparison of linear models including ridge regression and lasso regression with a non-linear random forest method, to find associations between brain imaging phenotypes and genotypes.

Our results show that lasso and random forest regression, but not ridge regression, identified a SNP in the APOE gene as the best performing variant. When compared across the whole genome, random forests produced a distinct list of selected SNPs, based on RMSE prediction performance, than the linear methods, which were highly similar to each other. Literature search and existing GWAS data showed that the top SNPs identified by random forests showed are all located in or near genes that have been previously associated with brain-related disorders, supporting the use of non-linear multi-variate GWAS methods to identify distinct genetic variants than those selected by conventional linear methods. Further in-depth study of the genes identified in this analysis may contribute to a better understanding of their association with brain function. 

Extending the analysis to the top 1,000 SNPs predicted by random forests, we observed a clustering of image features, showing that groups of variants, not colocated on the genome, tend to associate with similar brain regions or features. 

While reverse genotype prediction correctly picks up these correlations between the phenotypic traits, the corresponding correlations and shared effects between SNPs are currently ignored, since reverse genotype prediction approaches tend to learn prediction models for each SNP individually. Thus, a logical extension of our approach would be to use multi-task regression, i.e. to predict multiple SNPs simultaneously. However, this raises important computational challenges and it may be infeasible to predict SNPs simultaneously on a genome-wide scale. 

A disadvantage of using machine learning methods, in particular non-linear ones, is that the null distribution of the test statistic (RMSE) is unknown and the only way to quantify statistical significance is to compute permutation p-values, which was computationally infeasible across the whole genome. However, a more limited analysis on 876 SNPs showed that permutation p-values and random forest regression RMSE values, but not classification accuracies, showed a high degree of correlation. A possible solution could therefore be to learn a model to predict p-values from RMSE values from a suitable set of training SNPs, to be used to obtain approximate permutation p-values genome-wide.

Another limitation of the current study is that in the presence of highly correlated traits, the feature weights obtained by different methods are not necessarily robust. It would be interesting to investigate other measures of feature importance for random forest models, beyond the default ones based on gini importances, such as model-agnostic methods like permutation importance \cite{altmann2010permutation}.

Further future research could investigate additional non-linear machine learning methods such as neural networks, including deep neural networks for predicting genotypes using MRI recordings of the brain directly instead of extracted features \cite{lundervold2019overview}.  Moreover, since dementia is a progressive disorder, another interesting avenue to pursue would be to use longitudinal data.

\section*{Acknowledgements}
Data collection and sharing for this project was funded by the Alzheimer's Disease Neuroimaging Initiative (ADNI) (National Institutes of Health Grant U01 AG024904) and
DOD ADNI (Department of Defense award number W81XWH-12-2-0012). ADNI is funded
by the National Institute on Aging, the National Institute of Biomedical Imaging and
Bioengineering, and through generous contributions from the following: AbbVie, Alzheimer’s
Association; Alzheimer’s Drug Discovery Foundation; Araclon Biotech; BioClinica, Inc.;
Biogen; Bristol-Myers Squibb Company; CereSpir, Inc.; Cogstate; Eisai Inc.; Elan
Pharmaceuticals, Inc.; Eli Lilly and Company; EuroImmun; F. Hoffmann-La Roche Ltd and
its affiliated company Genentech, Inc.; Fujirebio; GE Healthcare; IXICO Ltd.; Janssen
Alzheimer Immunotherapy Research \& Development, LLC.; Johnson \& Johnson
Pharmaceutical Research \& Development LLC.; Lumosity; Lundbeck; Merck \& Co., Inc.;
Meso Scale Diagnostics, LLC.; NeuroRx Research; Neurotrack Technologies; Novartis
Pharmaceuticals Corporation; Pfizer Inc.; Piramal Imaging; Servier; Takeda Pharmaceutical
Company; and Transition Therapeutics. The Canadian Institutes of Health Research is
providing funds to support ADNI clinical sites in Canada. Private sector contributions are
facilitated by the Foundation for the National Institutes of Health (www.fnih.org). The grantee
organization is the Northern California Institute for Research and Education, and the study is
coordinated by the Alzheimer’s Therapeutic Research Institute at the University of Southern
California. ADNI data are disseminated by the Laboratory for Neuro Imaging at the University of Southern California.

\section*{Funding}
This work was supported in part by a grant from the Research Council of Norway (grant number 312045) to T.M, and in part by a grant from the Trond Mohn Research Foundation (grant number BFS2018TMT07) to A.S.L.

\bibliographystyle{unsrt}
\bibliography{refs}

\begin{thebibliography}{10}

\bibitem{shen2010whole}
Li~Shen, Sungeun Kim, Shannon~L Risacher, Kwangsik Nho, Shanker Swaminathan,
  John~D West, Tatiana Foroud, Nathan Pankratz, Jason~H Moore, Chantel~D Sloan,
  et~al.
\newblock Whole genome association study of brain-wide imaging phenotypes for
  identifying quantitative trait loci in {MCI} and {AD}: {A} study of the
  {ADNI} cohort.
\newblock {\em Neuroimage}, 53(3):1051--1063, 2010.

\bibitem{chauhan2015association}
Ganesh Chauhan, Hieab~HH Adams, Joshua~C Bis, Galit Weinstein, Lei Yu,
  Anna~Maria T{\"o}glhofer, Albert~Vernon Smith, Sven~J Van Der~Lee, Rebecca~F
  Gottesman, Russell Thomson, et~al.
\newblock Association of {Alzheimer}'s disease {GWAS} loci with {MRI} markers
  of brain aging.
\newblock {\em Neurobiology of aging}, 36(4):1765--e7, 2015.

\bibitem{bi2017genome}
Xuan Bi, Liuqing Yang, Tengfei Li, Baisong Wang, Hongtu Zhu, and Heping Zhang.
\newblock Genome-wide mediation analysis of psychiatric and cognitive traits
  through imaging phenotypes.
\newblock {\em Human brain mapping}, 38(8):4088--4097, 2017.

\bibitem{lu2017bayesian}
Zhao-Hua Lu, Zakaria Khondker, Joseph~G Ibrahim, Yue Wang, Hongtu Zhu,
  Alzheimer’s Disease~Neuroimaging Initiative, et~al.
\newblock Bayesian longitudinal low-rank regression models for imaging genetic
  data from longitudinal studies.
\newblock {\em NeuroImage}, 149:305--322, 2017.

\bibitem{liu2014cardiovascular}
Guiyou Liu, Lifen Yao, Jiafeng Liu, Yongshuai Jiang, Guoda Ma, Zugen Chen, Bin
  Zhao, Keshen Li, et~al.
\newblock Cardiovascular disease contributes to {Alzheimer}'s disease: evidence
  from large-scale genome-wide association studies.
\newblock {\em Neurobiology of aging}, 35(4):786--792, 2014.

\bibitem{potkin2009genome}
Steven~G Potkin, Jessica~A Turner, Guia Guffanti, Anita Lakatos, Federica
  Torri, David~B Keator, and Fabio Macciardi.
\newblock Genome-wide strategies for discovering genetic influences on
  cognition and cognitive disorders: methodological considerations.
\newblock {\em Cognitive neuropsychiatry}, 14(4-5):391--418, 2009.

\bibitem{wang2012identifying}
Hua Wang, Feiping Nie, Heng Huang, Sungeun Kim, Kwangsik Nho, Shannon~L
  Risacher, Andrew~J Saykin, Li~Shen, and Alzheimer's Disease~Neuroimaging
  Initiative.
\newblock Identifying quantitative trait loci via group-sparse multitask
  regression and feature selection: an imaging genetics study of the {ADNI}
  cohort.
\newblock {\em Bioinformatics}, 28(2):229--237, 2012.

\bibitem{wang2012phenotype}
Hua Wang, Feiping Nie, Heng Huang, Jingwen Yan, Sungeun Kim, Kwangsik Nho,
  Shannon~L Risacher, Andrew~J Saykin, Li~Shen, and Alzheimer's
  Disease~Neuroimaging Initiative.
\newblock From phenotype to genotype: an association study of longitudinal
  phenotypic markers to {Alzheimer}'s disease relevant {SNPs}.
\newblock {\em Bioinformatics}, 28(18):i619--i625, 2012.

\bibitem{huang2015fvgwas}
Meiyan Huang, Thomas Nichols, Chao Huang, Yang Yu, Zhaohua Lu, Rebecca~C
  Knickmeyer, Qianjin Feng, Hongtu Zhu, Alzheimer's Disease~Neuroimaging
  Initiative, et~al.
\newblock {FVGWAS}: Fast voxelwise genome wide association analysis of
  large-scale imaging genetic data.
\newblock {\em Neuroimage}, 118:613--627, 2015.

\bibitem{huang2019spatial}
Meiyan Huang, Chunyan Deng, Yuwei Yu, Tao Lian, Wei Yang, Qianjin Feng,
  Alzheimer's Disease~Neuroimaging Initiative, et~al.
\newblock Spatial correlations exploitation based on nonlocal voxel-wise {GWAS}
  for biomarker detection of ad.
\newblock {\em NeuroImage: Clinical}, 21:101642, 2019.

\bibitem{zhou2018brain}
Tao Zhou, Kim-Han Thung, Mingxia Liu, and Dinggang Shen.
\newblock Brain-wide genome-wide association study for {Alzheimer}'s disease
  via joint projection learning and sparse regression model.
\newblock {\em IEEE Transactions on Biomedical Engineering}, 66(1):165--175,
  2018.

\bibitem{hariri2006imaging}
Ahmad~R Hariri, Emily~M Drabant, and Daniel~R Weinberger.
\newblock Imaging genetics: perspectives from studies of genetically driven
  variation in serotonin function and corticolimbic affective processing.
\newblock {\em Biological psychiatry}, 59(10):888--897, 2006.

\bibitem{brun2009mapping}
Caroline~C Brun, Natasha Lepor{\'e}, Xavier Pennec, Agatha~D Lee, Marina
  Barysheva, Sarah~K Madsen, Christina Avedissian, Yi-Yu Chou, Greig~I
  De~Zubicaray, Katie~L McMahon, et~al.
\newblock Mapping the regional influence of genetics on brain structure
  variability—a tensor-based morphometry study.
\newblock {\em Neuroimage}, 48(1):37--49, 2009.

\bibitem{shen2007morphometric}
Li~Shen, Andrew~J Saykin, Moo~K Chung, and Heng Huang.
\newblock Morphometric analysis of hippocampal shape in mild cognitive
  impairment: An imaging genetics study.
\newblock In {\em 2007 IEEE 7th International Symposium on BioInformatics and
  BioEngineering}, pages 211--217. IEEE, 2007.

\bibitem{potkin2009hippocampal}
Steven~G Potkin, Guia Guffanti, Anita Lakatos, Jessica~A Turner, Frithjof
  Kruggel, James~H Fallon, Andrew~J Saykin, Alessandro Orro, Sara Lupoli, Erika
  Salvi, et~al.
\newblock Hippocampal atrophy as a quantitative trait in a genome-wide
  association study identifying novel susceptibility genes for {Alzheimer}'s
  disease.
\newblock {\em PloS one}, 4(8):e6501, 2009.

\bibitem{baranzini2009genome}
Sergio~E Baranzini, Joanne Wang, Rachel~A Gibson, Nicholas Galwey, Yvonne
  Naegelin, Frederik Barkhof, Ernst-Wilhelm Radue, Raija~LP Lindberg,
  Bernard~MG Uitdehaag, Michael~R Johnson, et~al.
\newblock Genome-wide association analysis of susceptibility and clinical
  phenotype in multiple sclerosis.
\newblock {\em Human molecular genetics}, 18(4):767--778, 2009.

\bibitem{ferreira2009multivariate}
Manuel~AR Ferreira and Shaun~M Purcell.
\newblock A multivariate test of association.
\newblock {\em Bioinformatics}, 25(1):132--133, 2009.

\bibitem{o2012multiphen}
Paul~F O’Reilly, Clive~J Hoggart, Yotsawat Pomyen, Federico~CF Calboli, Paul
  Elliott, Marjo-Riitta Jarvelin, and Lachlan~JM Coin.
\newblock Multiphen: joint model of multiple phenotypes can increase discovery
  in {GWAS}.
\newblock {\em PloS one}, 7(5):e34861, 2012.

\bibitem{malik2021high}
Muhammad~Ammar Malik, Adriaan-Alexander Ludl, and Tom Michoel.
\newblock High-dimensional multi-trait {GWAS} by reverse prediction of
  genotypes.
\newblock {\em arXiv preprint arXiv:2111.00108}, 2021.

\bibitem{banerjee2021tejaas}
Saikat Banerjee, Franco~L Simonetti, Kira~E Detrois, Anubhav Kaphle, Raktim
  Mitra, Rahul Nagial, and Johannes S{\"o}ding.
\newblock Tejaas: reverse regression increases power for detecting trans-eqtls.
\newblock {\em Genome biology}, 22(1):1--16, 2021.

\bibitem{fischl2012freesurfer}
Bruce Fischl.
\newblock {FreeSurfer}.
\newblock {\em Neuroimage}, 62(2):774--781, 2012.

\bibitem{edgington2007randomization}
Eugene Edgington and Patrick Onghena.
\newblock {\em Randomization tests}.
\newblock Chapman and Hall/CRC, 2007.

\bibitem{shadrin2021vertex}
Alexey~A Shadrin, Tobias Kaufmann, Dennis van~der Meer, Clare~E Palmer,
  Carolina Makowski, Robert Loughnan, Terry~L Jernigan, Tyler~M Seibert,
  Donald~J Hagler, Olav~B Smeland, et~al.
\newblock Vertex-wise multivariate genome-wide association study identifies 780
  unique genetic loci associated with cortical morphology.
\newblock {\em NeuroImage}, 244:118603, 2021.

\bibitem{squire2011cognitive}
Larry~R Squire and John~T Wixted.
\newblock The cognitive neuroscience of human memory since hm.
\newblock {\em Annual review of neuroscience}, 34:259--288, 2011.

\bibitem{amunts2005cytoarchitectonic}
Katrin Amunts, O~Kedo, M~Kindler, P~Pieperhoff, H~Mohlberg, NJ~Shah, U~Habel,
  F~Schneider, and K~Zilles.
\newblock Cytoarchitectonic mapping of the human amygdala, hippocampal region
  and entorhinal cortex: intersubject variability and probability maps.
\newblock {\em Anatomy and embryology}, 210(5):343--352, 2005.

\bibitem{poulin2011amygdala}
St{\'e}phane~P Poulin, Rebecca Dautoff, John~C Morris, Lisa~Feldman Barrett,
  Bradford~C Dickerson, Alzheimer's Disease~Neuroimaging Initiative, et~al.
\newblock Amygdala atrophy is prominent in early {Alzheimer}'s disease and
  relates to symptom severity.
\newblock {\em Psychiatry Research: Neuroimaging}, 194(1):7--13, 2011.

\bibitem{jack1998rate}
Clifford~R Jack, Ronald~C Petersen, Yuecheng Xu, Peter~C O'Brien, Glenn~E
  Smith, Robert~J Ivnik, Eric~G Tangalos, and Emre Kokmen.
\newblock Rate of medial temporal lobe atrophy in typical aging and
  {Alzheimer}'s disease.
\newblock {\em Neurology}, 51(4):993--999, 1998.

\bibitem{arnold1991topographical}
Steven~E Arnold, Bradley~T Hyman, Jill Flory, Antonio~R Damasio, and Gary~W
  Van~Hoesen.
\newblock The topographical and neuroanatomical distribution of neurofibrillary
  tangles and neuritic plaques in the cerebral cortex of patients with
  alzheimer's disease.
\newblock {\em Cerebral cortex}, 1(1):103--116, 1991.

\bibitem{khan2014molecular}
Usman~A Khan, Li~Liu, Frank~A Provenzano, Diego~E Berman, Caterina~P Profaci,
  Richard Sloan, Richard Mayeux, Karen~E Duff, and Scott~A Small.
\newblock Molecular drivers and cortical spread of lateral entorhinal cortex
  dysfunction in preclinical {Alzheimer}'s disease.
\newblock {\em Nature neuroscience}, 17(2):304--311, 2014.

\bibitem{nestor2008ventricular}
Sean~M Nestor, Raul Rupsingh, Michael Borrie, Matthew Smith, Vittorio
  Accomazzi, Jennie~L Wells, Jennifer Fogarty, Robert Bartha, and Alzheimer's
  Disease~Neuroimaging Initiative.
\newblock Ventricular enlargement as a possible measure of {Alzheimer}'s
  disease progression validated using the {Alzheimer}'s disease neuroimaging
  initiative database.
\newblock {\em Brain}, 131(9):2443--2454, 2008.

\bibitem{wright2000meta}
Ian~C Wright, Sophia Rabe-Hesketh, Peter~WR Woodruff, Anthony~S David, Robin~M
  Murray, and Edward~T Bullmore.
\newblock Meta-analysis of regional brain volumes in schizophrenia.
\newblock {\em American Journal of Psychiatry}, 157(1):16--25, 2000.

\bibitem{kempton2011structural}
Matthew~J Kempton, Zainab Salvador, Marcus~R Munaf{\`o}, John~R Geddes, Andrew
  Simmons, Sophia Frangou, and Steven~CR Williams.
\newblock Structural neuroimaging studies in major depressive disorder:
  meta-analysis and comparison with bipolar disorder.
\newblock {\em Archives of general psychiatry}, 68(7):675--690, 2011.

\bibitem{macarthur2017new}
Jacqueline MacArthur, Emily Bowler, Maria Cerezo, Laurent Gil, Peggy Hall, Emma
  Hastings, Heather Junkins, Aoife McMahon, Annalisa Milano, Joannella Morales,
  et~al.
\newblock The new nhgri-ebi catalog of published genome-wide association
  studies (gwas catalog).
\newblock {\em Nucleic acids research}, 45(D1):D896--D901, 2017.

\bibitem{farrer1997effects}
Lindsay~A Farrer, L~Adrienne Cupples, Jonathan~L Haines, Bradley Hyman,
  Walter~A Kukull, Richard Mayeux, Richard~H Myers, Margaret~A Pericak-Vance,
  Neil Risch, and Cornelia~M Van~Duijn.
\newblock Effects of age, sex, and ethnicity on the association between
  apolipoprotein {E} genotype and {Alzheimer} disease: a meta-analysis.
\newblock {\em Jama}, 278(16):1349--1356, 1997.

\bibitem{van2020understanding}
Dennis van~der Meer, Oleksandr Frei, Tobias Kaufmann, Alexey~A Shadrin, Anna
  Devor, Olav~B Smeland, Wesley~K Thompson, Chun~Chieh Fan, Dominic Holland,
  Lars~T Westlye, et~al.
\newblock Understanding the genetic determinants of the brain with mostest.
\newblock {\em Nature communications}, 11(1):1--9, 2020.

\bibitem{grasby2020genetic}
Katrina~L Grasby, Neda Jahanshad, Jodie~N Painter, Luc{\'\i}a Colodro-Conde,
  Janita Bralten, Derrek~P Hibar, Penelope~A Lind, Fabrizio Pizzagalli,
  Christopher~RK Ching, Mary Agnes~B McMahon, et~al.
\newblock The genetic architecture of the human cerebral cortex.
\newblock {\em Science}, 367(6484):eaay6690, 2020.

\bibitem{wang2020genome}
Hui Wang, Jingyun Yang, Julie~A Schneider, Philip~L De~Jager, David~A Bennett,
  and Hong-Yu Zhang.
\newblock Genome-wide interaction analysis of pathological hallmarks in
  {Alzheimer}'s disease.
\newblock {\em Neurobiology of aging}, 93:61--68, 2020.

\bibitem{harrington1991measurement}
Charles~R Harrington, Elizabeta~B Mukaetova-Ladinska, Richard Hills, Patricia~C
  Edwards, E~Montejo De~Garcini, Michael Novak, and CM1905817 Wischik.
\newblock Measurement of distinct immunochemical presentations of tau protein
  in {Alzheimer} disease.
\newblock {\em Proceedings of the National Academy of Sciences},
  88(13):5842--5846, 1991.

\bibitem{altmann2010permutation}
Andr{\'e} Altmann, Laura Tolo{\c{s}}i, Oliver Sander, and Thomas Lengauer.
\newblock Permutation importance: a corrected feature importance measure.
\newblock {\em Bioinformatics}, 26(10):1340--1347, 2010.

\bibitem{lundervold2019overview}
Alexander~Selvikv{\aa}g Lundervold and Arvid Lundervold.
\newblock An overview of deep learning in medical imaging focusing on {MRI}.
\newblock {\em Zeitschrift f{\"u}r Medizinische Physik}, 29(2):102--127, 2019.

\end{thebibliography}

\newpage
\begin{center}
  {\LARGE \textbf{Supplementary Information}}
\end{center}

\section*{Supplementary Figures}
\renewcommand\thefigure{S\arabic{figure}}

\setcounter{figure}{0}

\begin{figure*}[h!]
 \begin{center}
  \begin{minipage}[t]{0.80\linewidth}
  \textbf{A}\\
      \includegraphics[width=\linewidth]{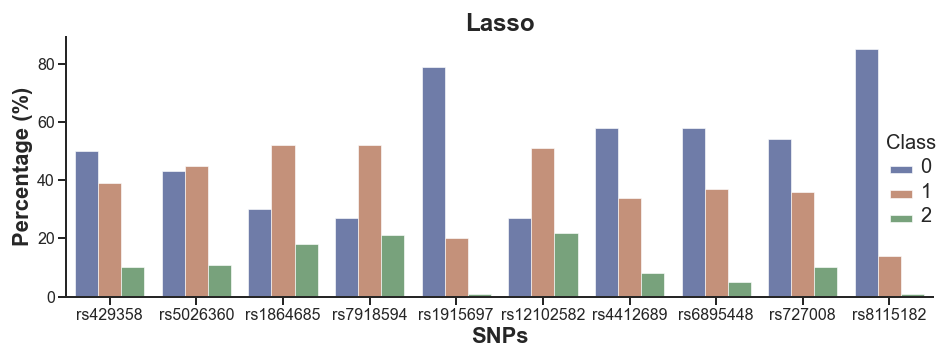}
  \end{minipage}
  \hfill
  \begin{minipage}[t]{0.80\linewidth}
  \textbf{B}\\
      \includegraphics[width=\linewidth]{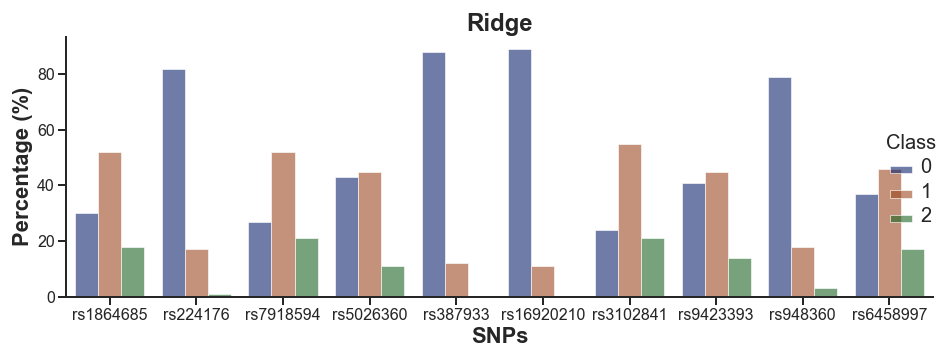}
  \end{minipage}
  \hfill
  
        \begin{minipage}[t]{0.80\linewidth}
        \textbf{C}\\
      \includegraphics[width=\linewidth]{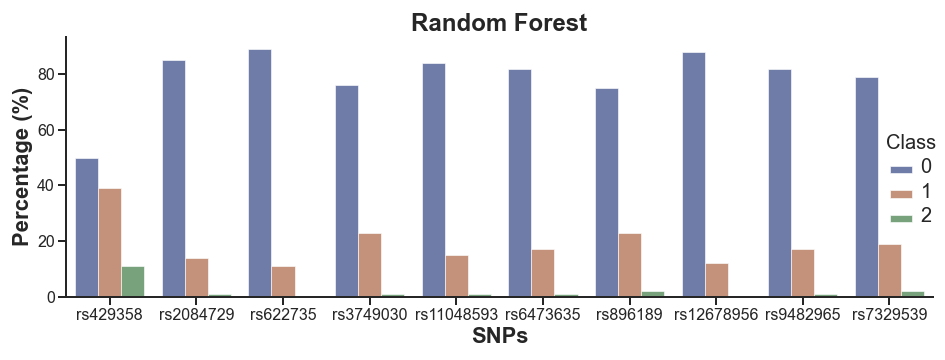}
  \end{minipage}
  \end{center}
  \caption{Class distribution for the top 10 SNPs identified by Random Forest, Lasso and Ridge regression.}
  \label{fig:S1}
\end{figure*}

\begin{figure}[t!]
\centering
\includegraphics[width=0.7\linewidth]{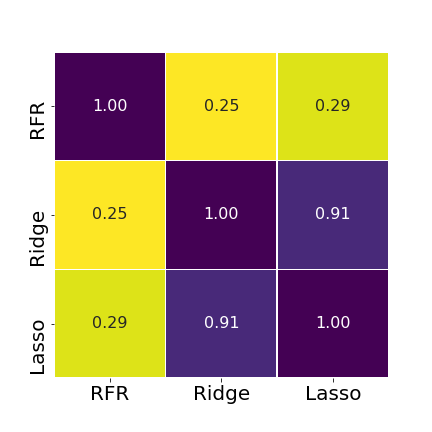}
\caption{Spearman correlation coefficients for RMSE values across the methods}
\label{fig:heatmap_methods}
\end{figure}

\begin{figure}[h!]
    \centering
    \includegraphics[width=0.80\linewidth]{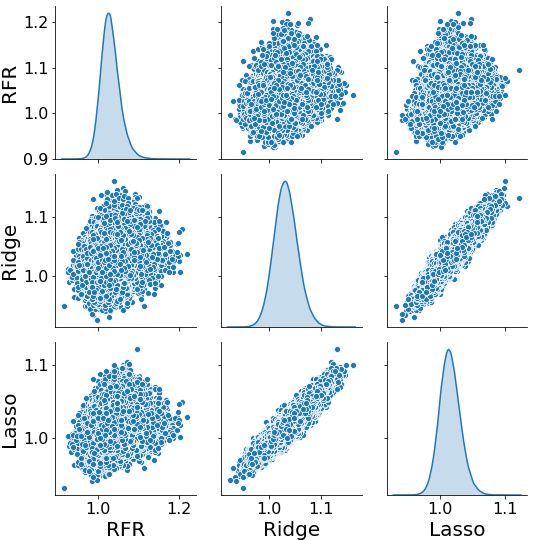}
    \caption{Pairplot showing the RMSE distribution and scatter plots between the methods, Random Forest Regression (RFR), Ridge Regression (RR), Lasso Regression (LR)}
    \label{fig:S3}
\end{figure}

\begin{figure}[h!]
    \centering
    \includegraphics[width=0.80\linewidth]{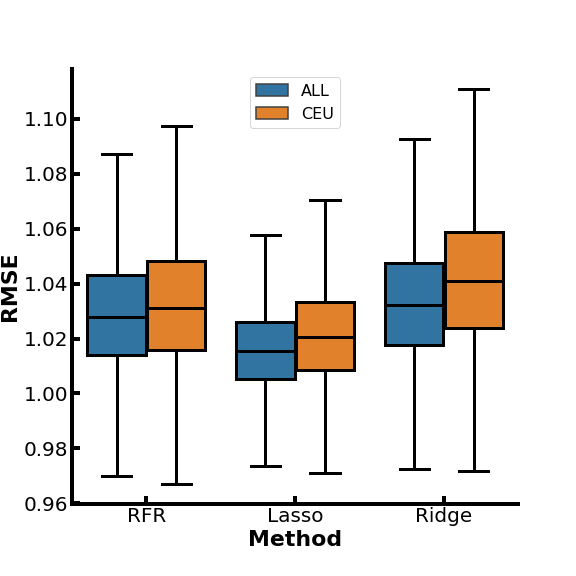}
    \caption{Boxplots showing RMSE distribution across all SNPs for Random Forest Regression (RFR), Lasso Regression and Ridge Regression, for all samples vs only Caucasian samples}
    \label{fig:S2}
\end{figure}

\end{document}